\documentstyle[prd,aps,preprint]{revtex}
\tightenlines
\input epsf.tex
\def\DESepsf(#1 width #2){\epsfxsize=#2 \epsfbox{#1}}

\begin{document}

\draft
\preprint{\vbox{
\hbox{UMD-PP-00-019}}}
\title{ Breaking Parity Symmetry Using Extra Dimensions}

\author{ R. N. Mohapatra$^1$\footnote{e-mail:rmohapat@physics.umd.edu}
and
A. P\'erez-Lorenzana$^{1,2}$\footnote{e-mail:aplorenz@Glue.umd.edu} }

\address{$^1$ Department of
Physics, University of Maryland, College Park, MD, 20742, USA\\
$^2$  Departamento de F\'\i sica, 
Centro de Investigaci\'on y de Estudios Avanzados del I.P.N.\\
Apdo. Post. 14-740, 07000, M\'exico, D.F., M\'exico.}
\date{September, 1999}
\maketitle
\begin{abstract}
{ We present a new way to break parity symmetry in left-right symmetric 
models using boundary conditions on the fields residing in the  fifth
dimension. We also discuss the connection between the limits on the size
of extra dimensions and the scale of right handed symmetry breaking
obtained from the analysis of neutrinoless double beta decay in the case
where the righthanded gauge symmetry is in the bulk.}.\\[1ex]
pacs: { 12.60.Cn; 11.30.Qc; 11.30.Er; 11.10.Kk}
 \end{abstract} 
 
\vskip0.5in

\section{Introduction}
The possible existence of hidden extra dimensions, motivated by
superstring
theories has provided a new way to look at many particle physics
phenomena such as unification of couplings, baryon and lepton
nonconservation\cite{many1,many2,lykken,many3,ddg,many,nath}
 and cosmological ones such as inflation,
baryogenesis\cite{cosmo}. Perhaps one of the most interesting
experimental fallouts of this line of research has come from the
realization that extra
dimensions almost as large as a millimeter could apparently be
hidden from many extremely precise measurements that exist in
particle physics leading to new searches for tests of gravitation at
submillimeter distances. Also exciting are the possibilities that
the concept of hidden space dimensions can be probed by collider
as well as other experiments in not too distant future. 

In this note, we point out another application of the extra dimensions.
We show that the use of boundary conditions on fields in the extra
dimensions provides a new way to break the parity symmetry of the
left-right symmetric models.  This idea is
analogous to that of supersymmetry breaking
by choosing different boundary conditions of various members of the
supermultiplets advocated long ago by Scherk and Schwarz\cite{ss}.

We show that in the case when the bulk field
used is fermionic,
starting with a completely parity symmetric Lagrangian in the brane
with positive mass terms for the Higgs fields, the contributions of the
bulk fermion fields leads to a radiative symmetry breaking of
parity symmetry. In this framework, the parity
breaking scale $v_R$ and the string scale, $M_{str}$ are of
the same order if we use the string scale as the cutoff in divergent
Feynman loops in the effective four dimensional field theory.
For the sake of comparison we point out that in the literature there
exist two ways to break left-right symmetry in  the context of
$SU(2)_L\times SU(2)_R\times U(1)_{B-L}$ models: one is to use different
masses for the Higgs fields $\chi_L,\chi_R$ thereby breaking parity
softly\cite{mp} and a second one is to keep the weak Lagrangian exactly
parity symmetric but looking for a parity asymmetric minimum which can
exist for a domain of the parameters in the Higgs potential\cite{senj}. 
In both these cases, the value of the parity breaking scale is an
arbitrary parameter. There is however a difference between these two
methods in that the second case could apriori lead to the domain wall
problem in the early universe, to cure which one may need additional
physics inputs\cite{mohsenj}. From the way the new mechanism suggested in
this article is implemented, it appears that the domain wall problem
should be absent in the present case. 

We then discuss the
process of neutrinoless double beta decay in the presence of extra
dimensions and show that the higher Kaluza-Klein modes of the righthanded
W-boson provide new contributions to this process. We compute these
contributions and obtain correlated limits on $m_{W_R}$ and the inverse
size of the extra dimensions. We find that due to already existing limits
on $R^{-1}$ from the considerations of the standard model
phenomena\cite{nath}, the limits on $W_R$ remain same as in the case
without the extra dimensions. Once the limits on the double beta lifetime
improve, one could expect more stringent limits on $W_R$ as well as
$R^{-1}$. This discussion applies regardless of how parity is broken.

\section{A toy model}

Let us start our discussion by considering a toy model with a discrete
symmetry. Consider a model with a $Z_2$ symmetry under which two scalar
fields $\chi_L$ and $\chi_R$ living in the 3+1 dimensional brane go into
each other. Let us assume that the brane is embedded in a 4+1 dimensional
spacetime. Consider a pair of bulk fields $\sigma_L\oplus \sigma_R$ which
also go into each other under the $Z_2$ symmetry. The action for this
system can be written as a sum of three terms:
\begin{eqnarray}
{\cal S}= {\cal S}_4(\chi_L, \chi_R)+{\cal S}_5(\sigma_L,\sigma_R) +{\cal
S}_{45}(\chi_L,\chi_R,\sigma_L, \sigma_R)
\end{eqnarray}
The detailed form of the four dimensional part of the action ${\cal S}_4$
is obvious; below we explicitly give the other terms: 
 \begin{eqnarray}
{\cal S}_5 &=& \int d^4x dy
\left[\partial^M\sigma^{\dagger}_L\partial_M\sigma_L + L\leftrightarrow R +
...\right]\\
\nonumber
{\cal S}_{45}&=&\int d^4x \left[\chi^{\dagger}_L\chi_L\sigma_L(y=0) +
L\leftrightarrow R \right]
\end{eqnarray}
Now comes the crucial point of our paper. In order to evaluate the ${\cal
S}_{45}$, we need to know the boundary conditions on the bulk fields
$\sigma_{L,R}$. Suppose we impose the boundary conditions as follows:
\begin{eqnarray}
\sigma_L(x, -y)=-\sigma_L(x, y)\\ \nonumber
\sigma_R(x, -y)=\sigma_R(x, y)
\end{eqnarray}
The fields can then be Fourier expanded in the interval $-\pi R \leq y
\leq +\pi R$ as follows:
\begin{eqnarray}
\sigma_L(x, y) = \frac{1}{\sqrt{\pi R}}\sum_{n} \sigma^{(n)}_L \sin
\frac{ny}{R} \\ \nonumber
\sigma_R(x, y) =\frac{1}{\sqrt{2\pi R}}\sigma^{(0)}_R+ \frac{1}{\sqrt{\pi
R}} \sum_{n\neq 0} \sigma^{(n)}_R \cos \frac{ny}{R}
\end{eqnarray}
Now we see that if the brane on which the $\chi$ fields live is located in
the bulk at the point $y=0$, then $\sigma_L(y=0)=0$. As a result the 3+1
dimensional brane Lagrangian is not $Z_2$ symmetric anymore. When we
compute the radiative corrections (cutting off the infinite integrals at
the string scale), we find $Z_2$ asymmetric contributions to the physical
parameters. In particular the self energy corrections to the $\chi$ mass
terms will be asymmetric with only the $\chi_R$ receiving contributions of
magnitude
\begin{equation}
m^2_{{1-loop}_R}\sim\sum_{KK}
 \frac{M^2_{str}}{M_{str}R}\approx M^2_{str}.
\end{equation}
where we have used the fact that the number of KK modes contributing to
the integral is roughly $M_{str}R$.
Important point
to note is that the bulk field flows in the loop and if we had a situation
where the bulk field was a fermion field, this contribution to mass would
be negative triggering break down of the discrete symmetry even if the
tree level terms in the brane were positive.
 In that case the string scale and the scale of $Z_2$ symmetry
breaking would necessarily be of same order. We will give an example of
this type in the
context of a realistic left-right symmetric model where the discrete
symmetry on the fields in the bulk arise quite naturally.

\section{Parity breaking in a left-right symmetric model}

These models which have been extensively discussed in the
literature are based on the gauge group $SU(2)_L\times
SU(2)_R\times U(1)_{B-L}$ with quark and lepton assignments into
left-right symmetric SU(2) doublets. We will not display this part of the
Lagrangian explicitly but rather focus on the part that is relevant to
our discussion of the symmetry breaking. We consider the nonsupersymmetric
version of the model and use the left-right Higgs doublets $\chi_L(2,1 +1)
\oplus \chi_R(1,2, +1)$ to break the gauge symmetry. We will also include
a bidoublet $\phi(2,2,0)$ to give mass to the charged fermions. We further
assume that
there is at least one bulk neutrino (denoted by $\nu_B$) which couples
to the brane fields. Also for simplicity we will consider a five
dimensional theory although this is not essential for the discussion of
symmetry breaking.

Let us now discuss the parity transformation of the Higgs fields and the
bulk field. We use the straightforward definition of parity under which
$\psi_L\rightarrow \psi_R$ where $\psi$ denotes a typical fermion field
(both in the bulk and the brane). The Higgs fields transform as
$\chi_L\leftrightarrow \chi_R$.

As in the case of the toy model, the action will have three parts and we
will use the same notation as in the case of the toy model. In the 3+1
dimensional brane Lagrangian, in addition to the usual gauge invariant
pieces, we will assume that the parity invariant mass term for
$\chi_{L,R}$ fields is small ($\sim m_W$ or smaller).

 Let us write down the relevant terms in the action:
\begin{eqnarray}
{\cal S}_{45}&=& \int d^4x \left[\kappa\bar{L}\chi_L \nu_{BR}(x, y=0) +
\kappa \bar{R}\chi_R \nu_{BL}(x, y=0)+ h.c.\right]\\ \nonumber
 {\cal S}_5 &=& \int d^4x dy
\bar{\nu}_B\Gamma^5\partial_5\nu_B
\end{eqnarray}
where $L^T=(\nu_{eL}, e_L)$ and $R^T=(\nu_{eR}, e_R)$

In order to break the $SU(2)_R$ gauge group, we need to give $<\chi^0_R>=
v_R$. In previous discussions of this\cite{senj}, one searches for a
domain of the coupling parameters in the Higgs potential
\begin{eqnarray}
V(\chi_L, \chi_R) = \mu^2 (\chi^{\dagger}_L\chi_L +\chi^{\dagger}_R\chi_R)
+\lambda_{+}(\chi^{\dagger}_L\chi_L +\chi^{\dagger}_R\chi_R)^2\\ \nonumber
+\lambda_{-}(\chi^{\dagger}_L\chi_L -\chi^{\dagger}_R\chi_R)^2
\end{eqnarray}
One finds that for $\lambda_{-}< 0$, the minimum of the above potential is
parity violating. What we will show is that using the new way we propose,
one can break parity symmetry even for $\lambda_{-}\geq 0$.

Crucial to implementing our mechanism of parity breaking are the boundary
conditions on the bulk field. We impose the following condition:
\begin{eqnarray}
\nu_{B}(x, -y)=-\gamma_5\nu_B(x, y)
\end{eqnarray}
This implies that 
\begin{eqnarray}
\nu_{BL}(x, -y) &=& +\nu_{BL}(x,+y) \\ \nonumber
\nu_{BR}(x, -y) &=& -\nu_{BR}(x,+y) 
\end{eqnarray}
This allows us to write the following Fourier expansions for the
$\nu_{BL,R}$ fields in the extra dimension $y$ which is compactified and
assumed to be in the domain $-\pi R\leq y \leq \pi R$ as in the previous
case.
\begin{eqnarray}
\nu_{BR}(x, y)&=&  \frac{1}{\sqrt{\pi R}}\sum_{n} \nu^{(n)}_{BR} 
\sin\frac{ny}{R} \\ \nonumber
\nu_{BL}(x, y) &=& \frac{1}{\sqrt{2\pi R}}\nu^{(0)}_{BL}
+ \frac{1}{\sqrt{\pi R}} \sum_{n\neq 0} \nu^{(n)}_{BL} \cos \frac{ny}{R}
\end{eqnarray}
Now we see that since the brane is located at the point $y=0$,
$\nu_{BR}(x, y=0)=0$ and therefore the $L_L\chi_L\nu_{B}$ coupling
vanishes. This induces the breaking of parity symmetry in the brane
Lagrangian.

One practical consequence of this explicit breaking (induced by the bulk)
is that the one loop self energy contribution from the $\nu_B$
intermediate state to the $\chi_L$ field vanishes whereas for the
$\chi_R$, we get
\begin{eqnarray}
m^2_{\chi_R}(1-loop)\sim - \frac{1}{M_{str}R}\sum_n \int
\frac{d^4k}{(2\pi)^4}
\frac{(k^2-p.k)}{k^2(k^2-2p.k+p^2 -\frac{n^2}{R^2})}
\end{eqnarray}
This integral has a quadratic dependence on the cut off which is
independent of the mass of the leptons running into the loop.  Therefore,
all the KK modes contribute roughly  with the same amount to the mass term. 
We estimate this to be $m^2_{\chi_R}\sim - a M^2_{str}$. Note that this
contribution to the scalar
mass is negative which therefore can trigger the breaking of parity
symmetry leading to the relation
\begin{eqnarray} v_R\sim M_{str}
\end{eqnarray}
Thus this way of parity breaking enables us to relate the otherwise free
parameter $v_R$ to more fundamental scales in the theory. Note that
this new negative contribution to the self mass of the $\chi_R$ is present
only in
the presence of the bulk neutrino. For instance, in its absence the only
self energy contribution to $\chi_R$ would come from the scalar self
couplings and would be positive.

 Admittedly the discussions given above have been very simpleminded with 
no pretense to rigor; however we believe that we have
pointed out a fundamentally new way to discuss breaking of discrete
symmetries. In fact it would be quite interesting to search for a model of
CP violation that uses this mechanism.

Let us close this section with some brief comments on some other
implications of our work.

\noindent{(i)} If we consider a five dimensional theory (or a theory with 
one large dimension and other dimensions with size $\sim M^{-1}_{str}$),
then the relation between the Planck mass and the string scale becomes:
\begin{eqnarray}
 M^2_{P\ell}\simeq M^3_{str} R
\end{eqnarray}
where R is the size of the extra dimension. Since gravity experiments
require that $R\leq $ 1 mm, this implies that $M_{str}\geq 10^8$ GeV.
This would then imply that there is a lower limit on the right handed
scale of $10^8$ GeV in our way of breaking right handed symmetry. Here we
have assumed that other extra dimensions have sizes of order
$M^{-1}_{str}$ or larger; otherwise the lower limit on the string scale
and hence the right handed scale becomes weaker. It is
important to note that existence of millimeter size extra dimensions are
compatible with the new way of parity breaking.

\noindent{(ii)} If we want to implement this way of parity breaking
in models where the righthanded symmetry is broken by $B-L=2$
triplets\cite{moha2} $\Delta_R(1,3 +2)\oplus \Delta_L(3,1,+2)$, then we
have to put these multiplets in the bulk. This requires for consistency
that $SU(2)_L\times SU(2)_R\times U(1)_{B-L}$ must also be in the bulk.
This automatically implies that the bulk size cannot be larger than a
TeV$^{-1}$\cite{nath}. In this case, processes such as neutrinoless double
beta decay receive new contributions and we discuss it in the next
section.

\noindent{(iii)} This model leads to the profile of
neutrino masses discussed in the Ref.\cite{mnp}. Otherwise there would be
an additional term that will mix the $\nu_L$'s with the bulk neutrinos and
that coupling has to be tuned down to an appropriate level to make the
scheme of Ref.\cite{mnp} work.

\noindent{(iv)} In this theory the universe never reaches
a symmetric parity invariant phase. As a result, there is no domain wall
problem as already noted. There may be additional cosmological
consequences of the bulk neutrinos, specially one might wonder about the 
impact of the heavy modes of the bulk neutrino. This depends
on the picture of inflation in such models and at this stage of the
development of the field, the answers are not clear.

\noindent{(v)} The bulk neutrinos couple to the standard model matter only
via the right handed Higgs fields, which being longitudinal modes of the
$W_R$ are very heavy. Therefore, it is not possible to get any constraints
on the bulk neutrino properties from known low energy electroweak data.

 Finally, we note that the gauge couplings also receive one loop 
parity asymmetric corrections which therefore make the gauge couplings
unequal. Their difference is connected to logarithm of the string scale
and is therefore small (of the order of the usual RGE corrections).

\section{Left-right symmetry in higher dimensions and neutrinoless double
beta decay}
In this section we discuss the case when the left-right gauge fields are
in the bulk as are the Higgs fields that break the righthanded gauge
symmetry. The fermions are in the brane and the usual seesaw mechanism
($\Delta_{R}(1,3 +2)$ breaking the righthanded symmetry) is used to get
small neutrino masses. Parity could be broken by the usual way using the
potential or by using the boundary conditions on Higgs fields propagating
in the bulk. In this case, there will be higher Kaluza-Klein modes of the
$W_R$ that will contribute to all processes where the right handed W used
to contribute. In particular, it will contribute to neutrinoless double
beta decay\cite{rnm} via the exchange of $W_R$ and the heavy righthanded
Majorana neutrino. For simplicity we will consider the case where masses
of the $W_R$ and the righthanded neutrino are same i.e. $M_{W_R}=M_{N_R}$.
In the presence of the KK models, the contribution to the
strength of the neutrinoless double beta amplitude can be written as:
\begin{eqnarray}
M_{\beta\beta}\simeq {G^2_{F}\over 2}
 \frac{M^4_{W_L}}{M^5_{W_R}} f(M_{W_R},
R)
\end{eqnarray}
where $R$ is the size of the extra dimension (we assume only one
extra dimension to be important). We get
\begin{eqnarray}
f(M_{W_R}, R) = \left[1+{2} \sum_n\frac{M^2_{W_R}R^2}{M^2_{W_R}R^2 +
n^2}\right]^2\\ \nonumber
=\left[\pi M_{W_R}R~ \coth(\pi M_{W_R}R)~\right]^2
\end{eqnarray}
The present experimental bounds from the $^{76}$Ge experiment\cite{baudis}
on $f$ can be obtained using crude estimates of the nuclear matrix
elements by $p^3_{Fermi}$ and the result is
\begin{eqnarray}
\frac{M^4_{W_L}}{M^5_{W_R}} f(M_{W_R},R)\leq 2\times 10^{-7} ~~GeV^{-1}
\end{eqnarray}
In Fig. 1 we plot the resulting correlated limits on $R^{-1}$ and
$M_{W_R}$ for the present limit on the double beta amplitude from
$^{76}$Ge and an anticipated improvement on it by one order of magnitude
in future. The regions to the right of the lines is allowed yielding lower
limits on $R^{-1}$ and $M_{W_R}$.
 
One can see the asymptotic effects on the limit as follows:
\noindent{(i)} $M_{W_R}R \ll 1$: In this case,
\begin{eqnarray}
f(M_{W_R}, R)\approx \left(1 + \frac{\pi^2 M^2_{W_R}R^2}{3}\right)^2
\end{eqnarray}
Note that this is consistent with decoupling theorem in the limit of
vanishing radius of the extra dimensions. We see that the KK modes do not
have any effect on the limit on $M_{W_R}$ without them. This is the large
$R^{-1}$ domain in the figure. In the opposite extreme, we have
\noindent{(ii)} $M_{W_R}R \gg 1$: In this case,
\begin{eqnarray}
f(M_{W_R}, R)\approx (\pi M_{W_R}R)^2
\end{eqnarray}
For this case to hold, we must have $M_{W_R}\gg R^{-1}$; but already
electroweak physics implies that $R^{-1}\gg 1-5$ TeV\cite{nath}. In this
case also there are no stronger limits since as soon as $M_{W_R}\gg $ TeV,
the coefficient of $f$ in Eq. (16) becomes very small and the double beta
limit is always satisfied. The general quality of the constraints for
arbitrary $M_{W_R}$ and $R$ are given in Fig. 1 from where it is clear
that due to already existing constraints for the $SU(2)_L$ case, the
constraints on $M_{W_R}$ are not very useful at this stage. If however the
limit on the double beta amplitude goes down by another order of
magnitude, we can start seeing more useful constraints(see the line to
the far right in Fig. 1). There are now
several experimental proposals such as GENIUS\cite{klap},
CUORE\cite{frank} and MOON\cite{ejiri} which have
the potential to give such limits. Further details on this question are
now under study.

In conclusion, we have presented a new way to break parity (and other
discrete symmetries) using extra dimensions\cite{jeeva,recai}. We
construct explicit
examples of left-right symmetric models where this way of parity breaking
is realized. We then discuss the effect of extra dimensions on
neutrinoless double beta decay and find a correlated bound involving both
the size of an extra dimension and the the $W_R$ scale.

\vskip1em

{\it Acknowledgements.}   The work of RNM is supported by a grant from the
National Science Foundation under grant number PHY-9802551. The work of
APL is supported in part by CONACyT (M\'exico). 



\begin{figure}
\centerline{
\epsfxsize=250pt
\epsfbox{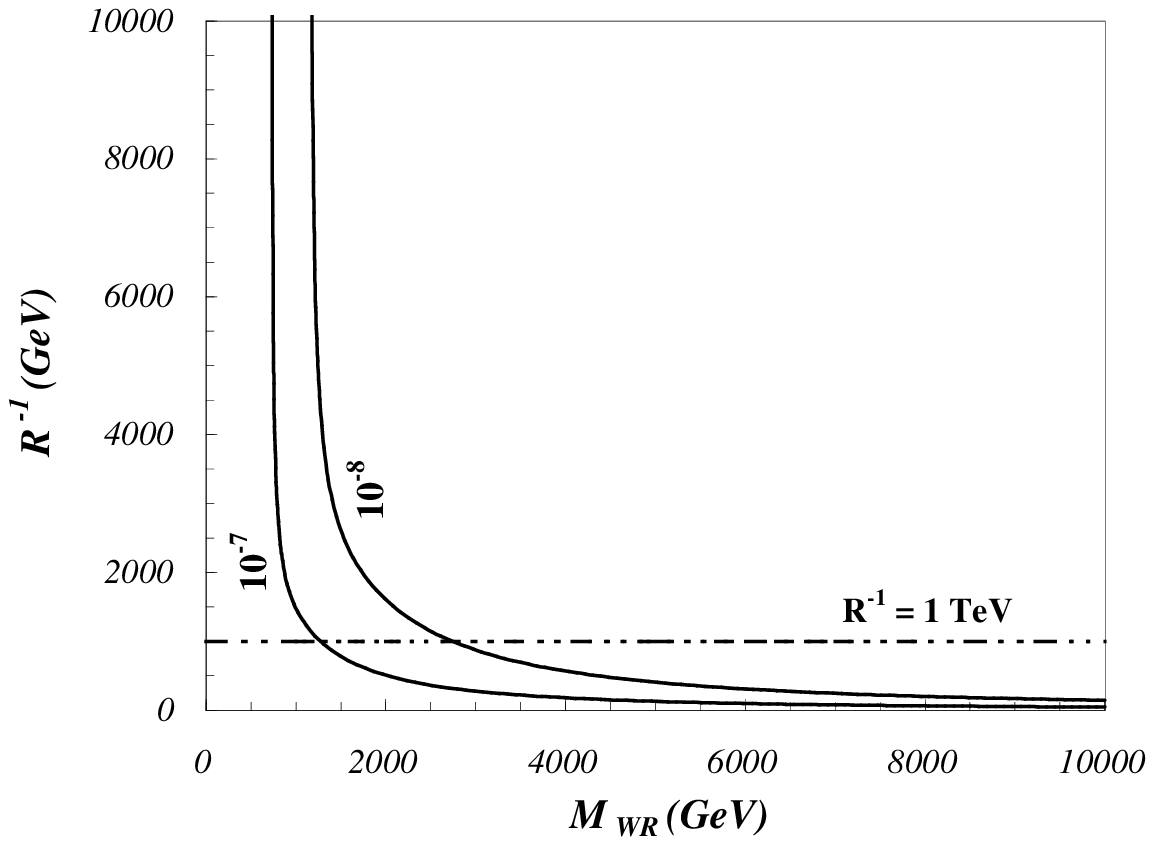}
}
\vskip1ex

\caption{Correlated limits for $M_{W_R}$ and $R$ from double beta decay
and electroweak physics. 
The labels  indicate the bounds for the amplitude  (solid
lines): $M^4_{W_L}  f(M_{W_R},R)/ 2 M^5_{W_R} $ at the current limit
($10^{-7}$ GeV$^{-1}$) 
and one order of magnitude lower; and for $R^{-1}= 1$ TeV
(dashed line). The
allowed region is above both limits.}
\end{figure}
\vskip2em

\end{document}